\documentclass[twocolumn,aps,pre,showpacs]{revtex4}

\usepackage{graphicx}

\usepackage{amsmath}
\usepackage{amssymb}

\begin{document}

\title{Semiclassical theory of chaotic quantum resonances}

\author{T.~Micklitz$^1$ and A.~Altland$^2$}

\affiliation{
$^1$Dahlem Center for Complex Quantum Systems and Institut 
f\"ur Theoretische Physik, Freie Universit\"at Berlin, 14195 Berlin, Germany\\
$^2$Institut f\"ur Theoretische Physik, Universit\"at zu K\"oln, Z\"ulpicher Str. 77, 50937 Cologne, Germany
}

\date{\today}


\begin{abstract}
  States supported by chaotic open quantum systems fall
  into two categories: a majority showing instantaneous
  ballistic decay, and a set of quantum resonances of
  classically vanishing support in phase space. We present a
 theory describing these structures within a unified
 semiclassical framework. 
   Emphasis is put on the quantum diffraction mechanism which 
 introduces an element of probability and is crucial for the 
 formation of resonances.
 Our main result are boundary conditions on the semiclassical 
 propagation along system trajectories. Depending on whether 
 the trajectory propagation time is shorter or longer than the 
 Ehrenfest time, these conditions describe deterministic escape, 
 or probabilistic quantum decay.
\end{abstract}

\maketitle

\section{Introduction} 
Quantum states populating 'open' chaotic
cavities decay to the outside environment and, thence, have the status
of resonances. In spite of the ubiquity of the general setup
--- open quantum chaos is realized in many of the devices currently
explored in mesoscopic physics, quantum optics, and cold atom physics
--- salient features of these resonances are not fully
understood. While the deep quantum regime (the Ehrenfest time, $t_E$,
marking the diffractive disintegration of minimal wave packages
shorter than classical escape times, $t_d$) appears to be under
reasonable control~\cite{Fyodorov}, it is the opposite, semiclassical
limit which  poses unsettled issues~\cite{marcel}.

Broadly speaking, the states populating an open cavity can be grouped
into two families: states evolving near classically and escaping
deterministically after a classical flight time, and a fraction
$\sim\exp(-t_E/t_d)$ of quantum resonances, whose probabilistic decay
is characterized by a finite imaginary offset $i\Gamma/2$ to the real
resonance energy $E$.  The most basic quantity characterizing the
statistics of resonances of complex energy $z=E+i{\Gamma\over 2}$ is
the resonance density $\rho(z)$.  Although the quantitative profile of
that quantity is not fully understood, the density appears to be
gapped against the real axis, $\Gamma=0$ (the existence of rare midgap
states notwithstanding~\cite{marcel,supersharp}.) The integrated
number of resonances at a given value of $E$ has been found to obey
the so-called fractal Weyl law, $\rho \propto \hbar^{-d_f}$, where
$d_f$ is a non-universal fractal exponent.

Previous work on the phenomenon includes the formulation of lower
bounds on the resonance
gap~\cite{gaspard,LuSridharZworski,nonnenmacher}, semiclassical
approaches based on short periodic orbits \textit{trapped} in the open
system~\cite{NovaesKeating}, a description in terms of non-unitarily
evolving Husimi functions~\cite{Keating}, phenomenology based on a
mixture of phase space dynamics and random matrix theory,
resp.~\cite{KoppSchomerus}, and numerical
analyses~\cite{marcel,Keating,LuSridharZworski,KoppSchomerus,Schomerus,bakersmap,shepelyansky}.
However, a unified theory of resonance formation in terms of first
principle semiclassical dynamics appears to be missing and the
formulation of such a theory is the subject of the present work.

Specifically, we will explore the quantum dynamics of states
concentrated on classical trajectories in terms of phase space Wigner
functions. Assuming globally hyperbolic classical dynamics we will
describe how quantum fluctuations operational on long trajectories
convert the deterministic classical escape of short trajectories into
probabilistic quantum decay.
 
Our analysis is organized in three conceptual steps. We first
introduce the phase space language used in the rest of the paper on a
one-dimensional toy model (section~\ref{sec1}). We then generalize to
the more complex setting of a higher dimensional cavity
(section~\ref{sec2}), and derive effective boundary conditions
determining the decay rates of the system. Finally,
(section~\ref{sec3}) we analyze these equations for both short and
long trajectories. We conclude in section~\ref{sec:summary}.

\section{one-dimensional toy model} 
\label{sec1}
\begin{figure}[h]
  \centering
 \centerline{\includegraphics[width=6.5cm]{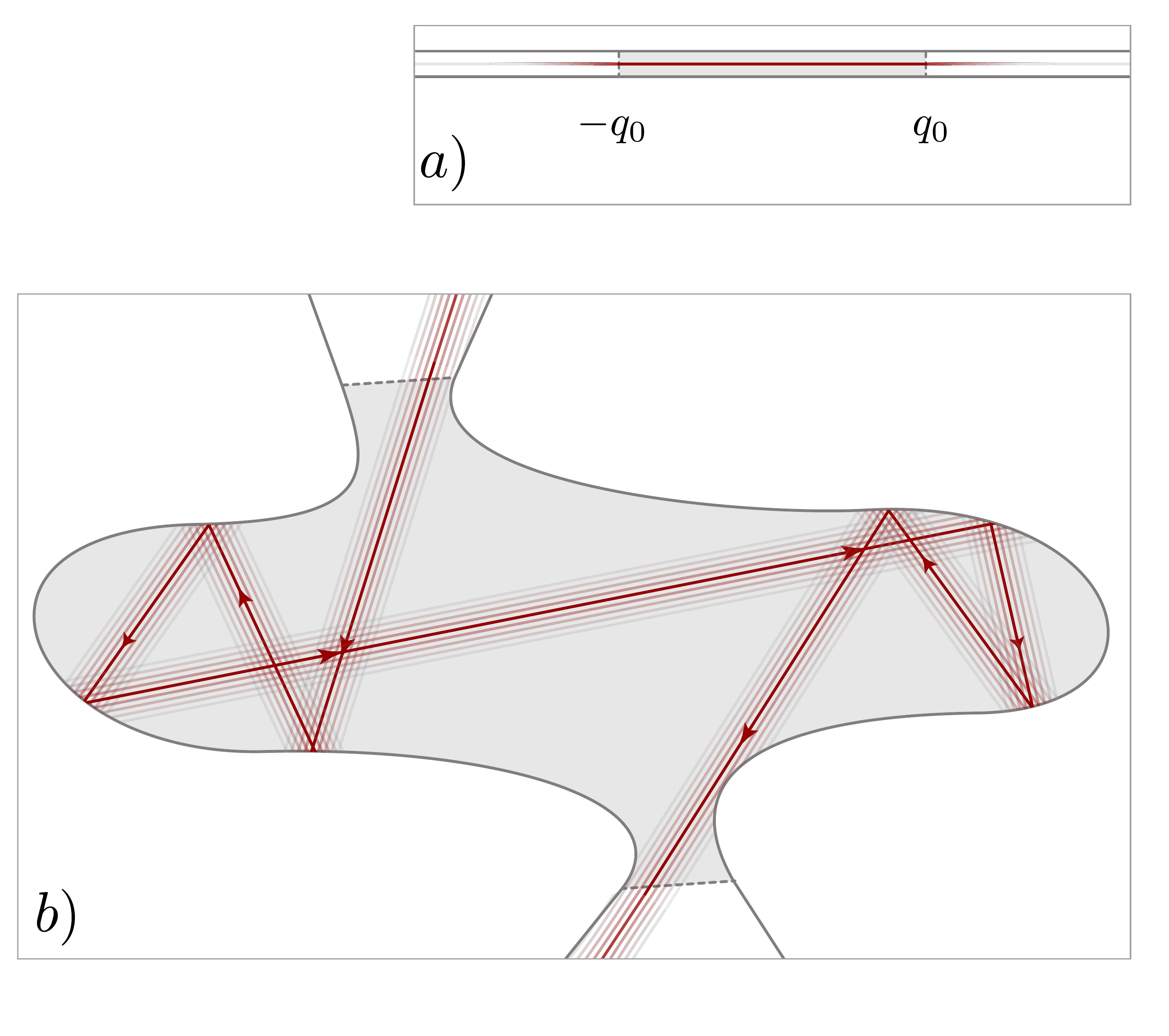}}
  \caption{Schematic illustration of the models discussed in the text: 
  (a) One-dimensional toy model consisting of a clean `cavity' (grey region)
  and `leads' connected to it (white regions). There is  no backscattering at the cavity/lead interfaces 
  and separation into cavity and leads is largely arbitrary.
    (b) Two-dimensional clean cavity with chaotic boundary scattering (grey region). 
    Contact to the `outside world' is due to fully transmitting openings connecting to the 
    reservoirs (white regions). Transitions from cavity to reservoirs are again smooth.
    \label{fig:0}
}
\end{figure}

Consider a one-dimensional 'cavity'
parameterized by the spatial coordinate $q\in [-q_0,q_0]$, while
coordinates to the right/left of $+q_0$/$-q_0$ define connecting
'leads' (see Fig.~\ref{fig:0}). We assume free intra-cavity particle dynamics,
$\hat{H}_0=\hat{p}^2/2m$ and, crucially, no backscattering barriers at
the cavity/lead interfaces.

Life times and energies of the resonant states supported by the system
may be calculated by matching solutions of the cavity Schr\"odinger
equation to outgoing boundary conditions~\cite{LandauQM,moiseyev},
i.e. by requiring that cavity wave functions $\psi(q)$ and their
derivatives smoothly connect to vacuum wave functions
$\varphi_\pm(q)\equiv a_\pm \exp( \pm i q \kappa )$ at the right/left interface.  
Here, $a_\pm=\mathrm{const.}$, and $\kappa=k -
i{k_\Gamma\over 2}$ is a complex wave vector whose real- and
imaginary-part define the energy $\hbar k=(2m E)^{1/2}$ and life time
$\hbar k_\Gamma\equiv \Gamma/v$ of resonant states, resp., where
$v=\hbar k/m$.  The divergence of the reference states at spatial
infinity $q\to \pm \infty$, is a formal means~\cite{LandauQM} to the
fixation of decay rates, as exemplified below.

For the intra-cavity wave function we make an ansatz
$\psi(q)
=\sum_{\sigma=\pm}ae^{\sigma(i\phi(q) +k_\Gamma vt(q)/2)}$ in terms
of left- and right-propagating partial amplitudes 
where $a=\mathrm{const.}$, and the somewhat unconventional denotation
$\phi(q)=kq$ and $t(q)=q/v$, for the real and imaginary contribution
to the phase, resp., will be motivated shortly. 
With this choice, the boundary conditions obtained by matching wave
functions and their derivatives at the left and right interface 
reduce to the single algebraic equation, 
\begin{align}
\label{bc}
e^{-(i2\phi(q_0)+k_\Gamma vt(q_0) )}
&=
{\partial_q\phi - k - i {k_\Gamma\over 2} \left( v \partial_q t - 1 \right)
\over 
\partial_q\phi + k - i {k_\Gamma\over 2} \left( v \partial_q t + 1 \right) }|_{q_0}.
\end{align}

Before evaluating this equation, let us translate
from the language of wave functions to a phase-space formulation. 
To this end, we introduce 
the Wigner function 
$W(q,p)=\int (d a)\, e^{-{ipa\over \hbar}} \bar \psi(q-{a\over 2})
\psi(q+{a\over 2})$, where 
$(da)= da/(2\pi\hbar)$. For our specific system,
\begin{align}
\label{toymodelWF}
W(q,p) 
&= \sum_{\sigma=\pm} a^2 \delta(p-\sigma p(E)) e^{\sigma  k_\Gamma v t(q)} + ...,
\end{align} 
where, $\sigma=\pm$ labels the Wigner transform of the left and right
moving components, resp., and the ellipses denote rapidly oscillating
interference contributions. In discarding 
the latter,  we loose track of the global phase of the wave
function, while the information on amplitudes and phase
\textit{derivatives} necessary to evaluate boundary conditions is retained.
Indeed, it is straightforward to check that 
\begin{align}
  \label{eq:4}
   a^2 e^{ \sigma k_\Gamma vt(q)} 
  =|\psi_\sigma(q)|^2
  &=\int (dp)\, W_\sigma(q,p),\cr
  \sigma \hbar \partial_q \phi(q) 
  &= {\int (dp)\, pW_\sigma(q,p)\over \int (dp)\, W_\sigma(q,p)}.
\end{align}
For the simple $1d$ system, the linear dependence $\phi_q =k q$
implies $\partial_q\phi=k$ so that \eqref{bc}
reduces to
\begin{align}
\label{bc1d}
{8E\over \Gamma} e^{-\Gamma t(q_0)/\hbar }
= 1 - v\partial_q t(q)|_{q=q_0}    
 \end{align} where we have 'fixed a gauge'  $e^{-2i\phi(q_0)}=i$
for the arbitrary phase of the wave function
 and neglected contributions $k_\Gamma/k\ll1$.

For the toy model at hand,  $v\partial_q t(q)= 1$, which means that 
 the right hand side of \eqref{bc1d} vanishes, 
 and $\Gamma\to \infty$ is the only consistent solution. This
 reflects the fact that a wave function will 'decay' with probability
 unity upon passing the reflectionless boundaries of the system. 
We next discuss how the situation changes upon generalization to a
higher dimensional system with chaotic 
dynamics.

\section{Chaotic cavity} 
\label{sec2}
We consider a two-dimensional cavity with
ballistic Hamiltonian $\hat H=\hat p^2/2m$ and chaotic boundary
scattering. The cavity is open such that after an average time $t_d$,
much shorter than any of the relevant quantum time scales,
trajectories escape through one or several reflectionless openings.
We define the Wigner function of the system's resonance states by
obvious generalization of Eq. \eqref{toymodelWF}, i.e.
$W(\mathbf{q},\mathbf{p})=\int (d^2a)\,e^{-i{\mathbf{p}\cdot
    \mathbf{a}\over \hbar}} \bar \psi(\mathbf{q}-{\mathbf{a}\over 2})
\psi(\mathbf{q}+{\mathbf{a}\over 2})$. To obtain the intra cavity
evolution equations of $W$, one adds and subtracts the Schr\"odinger
equations of the resonances $\psi$ and $\bar \psi$ to
obtain~\cite{textbook}
\begin{align}
\label{eq:2}
 \left[ H \stackrel{\ast}{,} W\right]_+ 
=2E\,  W, \quad
 \left[ H \stackrel{\ast}{,} W\right]_- 
= - i \Gamma\,  W.
\end{align}
Here, $H=p^2/2m$ is the Hamilton function and $\left[ A\,
  \stackrel{\ast}{,} B \right]_\mp=A\ast B\mp B\ast A$ where the Moyal
product of phase space functions $A=A(\mathbf{q},\mathbf{p})$ is given
by~\cite{MoyalProduct} $ A\ast B = A B + {i\hbar\over 2}\{A,B\} +
\mathcal{O}(\hbar^2), $ and $\{\,,\,\}$ is the Poisson bracket.

We next consider the vicinity of an exceptionally long trajectory
$\gamma_0$ spending time $T\gg t_d$ inside the cavity. For completeness we note that long
trajectories in open systems are found with low probability
$\sim\exp(-T/t_d)$. They typically form in the phase space neighborhood
of strange repellers realized through periodic orbits trapped
in the interior of the cavity (see below for further comments on this aspect).   
Assuming global
hyperbolicity of the dynamics, we introduce a trajectory coordinate,
$q\in [-vT/2,vT/2]$, a conjugate momentum $p=p(H)=(2mH)^{1/2}$
transverse to the shell of conserved energy, and a pair $u,s$ of
locally unstable and stable coordinates.  In the asymptotic
neighborhood of $\gamma_0$, the Hamiltonian can then be approximated
as $H\simeq H_0 = {p^2\over 2m} + \lambda us$, where $\lambda$ is a
Lyapunov exponent. The corresponding dynamics is generated by $[H_0
\stackrel{\ast}{,}\, . \, ]_-= i\hbar \{H_0,\;\}\equiv -i\hbar {\cal
  L}$, where the Liouvillian
\begin{align}
\label{dync}
{\cal L} &= v \partial_q + \lambda (u\partial_u - s\partial_s)
\end{align} 
describes propagation in the direction of $\gamma_0$, and exponential
expansion/contraction in the $u/s$ coordinate. Nonlinear corrections
to $H_0$ can be described as $H=H_0+V$, where $V=V(u,s)$ is a
polynomial of degree $> 2$ in the variables $u,s$. The corresponding
modification of the dynamics, $[V \stackrel{\ast}{,}\, . \, ]_-\equiv
-i\hbar (\Delta {\cal L}+{\cal Q})$,
comprises a weak alteration of the
classical Liouvillian,  $\Delta {\cal L}$, and a quantum generator
\begin{align}
\label{dynq}
{\cal Q} &= \sum_{n+m>1} c_{nm}\hbar^{n+m} \partial_u^m\partial_s^n,
\end{align} 
where $c_{nm}=c_{nm}(q,u,s)$ are coefficient functions whose
detailed profile will not be of much importance throughout. Although
both contributions are nominally small in $u,s$, the
quantum generator ${\cal Q}$, will be seen to have a
regularizing effect on classical singularities~\cite{ZurekPaz}, which
will ultimately shape the profile of the
resonance density.

\subsection{Life time in a chaotic cavity} 
Close to the trajectory, the
first of Eqs.~\eqref{eq:2}, $\left[ H \stackrel{\ast}{,}\, W
\right]_+\simeq \left[ H_0 \stackrel{\ast}{,}\, W \right]_+\simeq 2
(p^2/2m) W = 2E W$ simply describes the on-shell fixation $p\simeq
(2mE)^{1/2}$. Turning to the second equation,
\begin{align}
\label{qeq}
\hbar\left( {\cal L} + \Delta {\cal L} + {\cal Q} \right) 
W(q,u,s) = \Gamma W(q,u,s),
\end{align} 
we first discuss the linear approximation, $\Delta{\cal L}, {\cal
  Q}=0$, before including the correction terms in a second step. 

For $\Delta{\cal L}, {\cal Q}=0$, \eqref{qeq} 
becomes a first order differential equation which is solved in terms of a left-
and a right-moving contribution 
\begin{align}
\label{0QCSol}
W(q,p,u,s) 
&= \sum_{\sigma=\pm} a^2 \delta(p-\sigma p(E)) e^{\sigma k_\Gamma v t(q,u,s)},
\end{align}
structurally similar to Eq. \eqref{toymodelWF}. Here, $t(q,u,s)$ are
effective parameter functions generalizing $t(q)$ of the toy model and
evolving uniformly along the trajectories $\gamma\equiv
\gamma_\mathbf{x}$ piercing the phase-space point $\mathbf{x}\equiv
(q,u,s)$, ${\cal L} t(\mathbf{x})=1$. To solve this (partial first
order differential) equation, we consider its characteristics,
i.e. the trajectory $\gamma_\mathbf{x}$. On $\gamma_\mathbf{x}$, the
equation assumes the form $d_\tau t(\mathbf{x}(\tau))=1$, where $q(\tau)=q
+v_F \tau$,  
$u(\tau) = ue^{\lambda \tau}$,  $s(\tau)=s e^{-\lambda \tau}$ and
$(q,u,s)$ are starting values of the evolution. We solve the
characteristic equation as 
 $t(\tau)=\tau+t^0$, where $\tau$ increases uniformly
until $\gamma_{\mathbf{x}(\tau)}$ hits the
effective boundaries of the problem, and $t^0$ is a freely adjustable parameter. 

 \begin{figure}[h]
  \centering
 \centerline{\includegraphics[width=8.5cm]{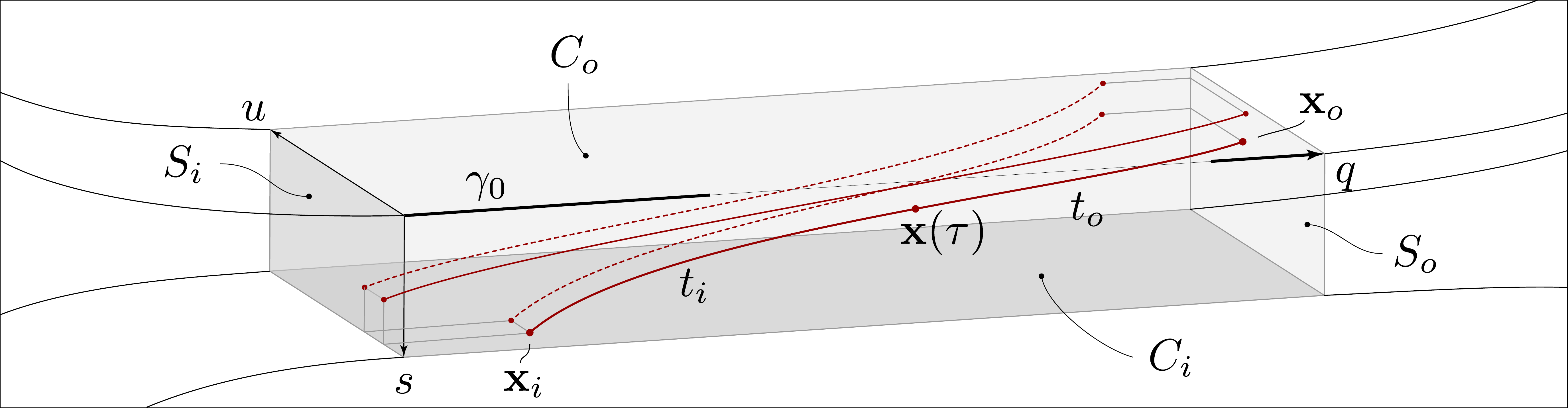}}
  \caption{
 On the definition of the different time parameters relevant to the
 trajectory dynamics. A trajectory $\gamma_0$ (indicated by a straight
 line along the upper front corner of the box) enters/exists the
 system through phase space interfaces $S_{i/o}$. Spanning the
 neighborhood of $\gamma_0$ by a stable/unstable coordinate $u$/$s$, nearby 
 trajectories contract towards/depart from $\gamma_0$
 in the respective coordinate directions. The exit out of the cavity
 neighborhood of $\gamma_0$ then is through the system interface $S_0$
 (solid trajectories), or through the 'internal interface', $C_o$
 (dashed trajectories.) A similar distinction applies to the entry
 points. Depending on the entry/exit variants, each trajectory
 neighboring $\gamma_0$ gets assigned entry/exit points
 $\mathbf{x}_{i/o}$, a time like progression
 parameter $\tau$, and a phase space parameterization $\mathbf{x}(\tau)$. 
\label{fig:1a}
}
\end{figure}

\begin{figure}[t!]
  \centering
 \centerline{\includegraphics[width=8.5cm]{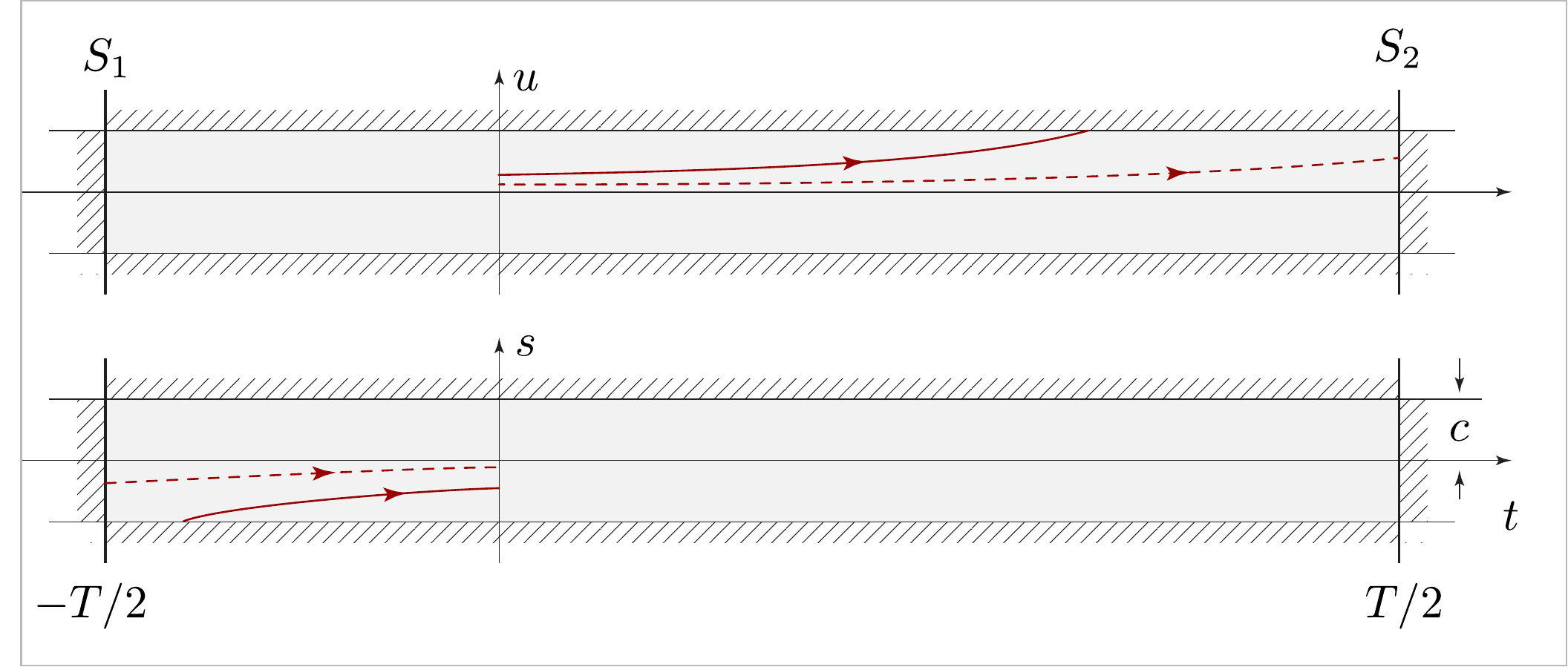}}
  \caption{
  The information of Fig.~\ref{fig:1a} collapsed to the
  two-dimensional sections spanned by the stable and the unstable
  coordinate, and the trajectory parameter, $q$, resp. A phase-space point $(q,u,s)$ in the vicinity of $\gamma_0$ 
 propagates along a unique classical trajectory, $\gamma$. It will exit the cavity 
 either through the interface $S_o$ or within the cavity
 through the surface $C_o$.
Similarly, the union of the left vacuum interface $S_i$ together with the manifold
$C_i$ 
defines the incoming interface. 
\label{fig:1}
}
\end{figure}

To understand the role of the boundaries, note
that $\gamma_{\mathbf{x}}$ will leave the cavity either through a
physical interface, $S_o$ along with $\gamma_0=\gamma_{(q,0,0)}$
(cf. Fig. \ref{fig:1}), or it will depart from $\gamma_0$ up to some
classical threshold $u\simeq c$ within the cavity (solid line). We
assume that points separated from $\gamma_0$ by scales $\sim c$ have
become generic and will exit in the classical, and hence negligibly
short time $\simeq t_d$. The union $I_o\equiv S_o\cup C_o$ of $S_o$
and the surface $C_o\equiv \{u=c,s,q\}$ then defines the effective
`outgoing interface' of our problem. Similarly, the union $I_i\equiv
S_i\cup C_i$ of the left vacuum interface $S_i$ and the surface
$C_i=\{u,s=c,q\}$ defines the incoming interface. The traveling phase space point
$\mathbf{x}(\tau)=(q,u,s)(\tau)$ hits the exit interface $I_o$, at the
smaller of two times, $\tau=t_o=t_o(q,u)=\mathrm{min}({T\over 2}-{q\over
  v},{1\over \lambda}\ln({c\over |u|}))$, depending on whether $S_o$
or $C_o$ is the terminal.  Likewise, $\gamma_\mathbf{x}$ has entered
the cavity through $I_i$ at a large negative time $\tau=-t_i=-t_i(q,s)
=-\mathrm{min}({T\over 2}+{q\over v},{1\over \lambda}\ln({c\over
  |s|}))$. Fixing the free parameter $t^0$ such that the temporal
range of the trajectory is symmetric around zero, 
$t(\tau=t_i)=-t(\tau=t_o)$, we find that the
solution to Eq. (\ref{0QCSol}) is governed by the function
$t(\mathbf{x})={1\over 2}(t_i(q,s)-t_o(q,u))$ while $T_\gamma\equiv
t_i+t_o$ is the intra cavity flight time of $\gamma$.
Notice that for points $\mathbf{x}\in I_o$ at the exit interface,
$t_o(\mathbf{x})=0$, meaning that
$t(\mathbf{x})=T_{\gamma}/2$ attains its maximal value. 

Finally, the boundary conditions Eq. \eqref{bc} are generalized by
replacing the one-dimensional variable $t(q)$ by 
$t(q,u,s)$, and the 
derivative $v\partial_q$ by $\mathcal{L}$,
i.e. a derivative acting in the direction of the Hamiltonian
flow~\cite{footnoteBC}. The generalization of Eq. \eqref{bc} thence
reads
\begin{align}
\label{gamma}
 {8E \over \Gamma} e^{-\Gamma t/\hbar} 
 = 1 - {\cal L} t,
\end{align} 
where $t=t(\mathbf{x})=T_\gamma/2$, and $\mathbf{x}\in I_o$ is on the
exit interface. 
Eq. \eqref{gamma} is a principal result of the present
paper.  In the following we discuss its implications for different
types of trajectories.

\section{Discussion of the results} 
\label{sec3}

\subsection{Short trajectories} 
To start with, we consider
trajectories $\gamma$ which are exceptionally long-lived, $T\equiv T_\gamma\gg
t_d$, yet short in comparison to the scale 
$t_E\sim{1\over \lambda} \ln(c^2/\hbar)$~\cite{ehrenfest} where quantum
uncertainty leads to the disintegration of semiclassically propagating
wave packets. 
As long as $T< t_E$,
the quantum generator
$\mathcal{Q}$ does not modify the dynamics in essential ways --- a
statement to be verified below --- and the same goes for the generator
of weak classical corrections, $\Delta \mathcal{L}$. We may thus take
the boundary condition \eqref{gamma} at face value, and conclude that
due to the homogeneous evolution 
$\mathcal{L} t=1$, the right hand 
side of the equation vanishes. As with the toy model, this implies a diverging decay rate,
$\Gamma\to \infty$. In this divergence reflects the fact that wave packages 
travelling on classically short trajectories leave the cavity with
certainty at the exit point.

\subsection{Long trajectories}
We now turn to  the case of long
trajectories, $T>t_E$. 
For asymptotically long trajectories, $T\to
\infty$, the classical shrinkage
$s=s(t)\sim c\exp(-\lambda t)$  would lead to singularities in the
function $t_i(q,s)$ and, ultimately, in the Wigner function. 
Within the present formalism, these singularities
are regularized on time scales larger than $t_E$, where
$s\lesssim s(t_E)\sim \hbar$ has shrunk down to quantum scales. On
these scales, the quantum generator $\mathcal{Q}$
of Eq. \eqref{dynq} is no longer small in comparison to the classical
generator $\mathcal{L}$ (while the correction $\Delta
\mathcal{L}$ to the classical flow continues to be largely
irrelevant.) The ensuing modifications of the dynamics can be described in 
various ways (cf.
Ref.~\cite{AndreevBilliards} for a treatment tailored to the formalism
applied here), the invariable conclusion being
that the shrinkage of classically evolving variables gets cut off by quantum
fluctuations. Technically, this conclusion rests on the observation that in the
evolution equation for the variable $s$ the higher order derivatives
$\sim \partial_s^{n>1}$ present in the quantum generator \eqref{dynq},
build up 'pressure' counteracting the classical contraction. This is
seen in explicit terms in the Fourier/Laplace representation of the
evolution equation, where these derivatives assume the form of
algebraic factors, cutting the logarithmic 'ultraviolet' singularities
of the classical equation. Referring to the appendix for
more details, we note that to leading semiclassical accuracy functions 
which in the classical theory evolve as
$f(q,|u|,|s|)$ get replaced by 
$f(q,|u|+\hbar/c,|s|+\hbar/c)$. Here, $c$ is symbolic notation for
classical ($\hbar$-independent) functions over which we have no
explicit control, and the substitution $|u|\to |u|+\hbar$ becomes
effectual in the large negative time asymptotics of a trajectory,
where $u$ rather than $s$ scales to small values.  

To understand the consequences of this regularization mechanism,
consider the trajectory time parameter, $t=t_i/2$ at the exit point of
$\gamma$. Now notice that $t_i(q,|s|) \to
\min(T/2+q/v,\lambda^{-1}\ln(|s|+\hbar/c))=\lambda^{-1}\ln(|s|+\hbar/c))\simeq
t_E$, where we used that, $T> t_E$. The crucial observation here is
that the regularization effectively truncates the in-time function
$t_i$ at values $t_E$. As a consequence, the interface derivative
$\mathcal{L} t={1\over2}\mathcal{L} (t_i-t_o)=1/2$ reduces to one half
of the value before quantum regularization. Substitution of this value
into Eq. \eqref{gamma} shows that the quantum theory admits finite
values of the decay constant, determined by
\begin{align}
  \label{eq:1}
 {\Gamma_0\over 2} ={\hbar \over t_E} W\left({8Et _E\over
    \hbar}\right)= {\hbar \over t_E} \left(
  \log\left({8E t_E \over \hbar}\right)+\dots\right),
\end{align}
where $W$ is the Lambert function and ellipses denote subleading
double-`$\log$' contributions. Eq.~\eqref{eq:1} states the decay rate
in terms of the Ehrenfest time in combination with non-universal short
time cutoff $\hbar/E$. However, in the semiclassical limit, $\hbar \to
0$, the dependence on $E$ drops out, and we are left with the
asymptote $\Gamma_0\sim \hbar \lambda$. Before commenting on this
result, we note that the appearance of a finite decay rate within our
present formalism follows from the fact that, by Heisenberg
uncertainty, quantum mechanics is not capable of resolving the phase
space fine structures pertaining to the evolution of long trajectories
$T_\gamma>t_E$. Each such trajectory should, rather, be thought of as
a distribution defined by the union of trajectories with uncertainty
$\sim \hbar$ in their phase space coordinates. At a given instance of
time, a fraction of this distribution escapes, as described by the
rate $\Gamma_0$.

 \subsection{Effective decay rate} 
 Our above analysis was oversimplifying in that it treated escape from
 an isolated long trajectory, $\gamma_0$, as tantamount to escape into
 the lead vacuum. This picture ignores the fact that escaping
 trajectories may get 'folded back' into the repeller domain
 supporting $\gamma_0$, and thence be trapped again. 
A statistical theory accounting for the renormalization of the decay
of an initial distribution centered around an isolated trajectory by
the complex structure of the embedding repeller structure has been
developed in Refs.~\cite{gaspard,jensen}. The result of that analysis
is an effective renormalization of decay rates as
$\lambda\to\lambda(1-d)$, where the factor $(1-d)$ effectively measures the  
fraction of
 trajectories managing to escape the repeller and $d$ is
 the fractal (information) dimension of the
 latter~\cite{gaspard,ott}. The
 ensuing effective rate, $\Gamma_0\to \Gamma\equiv \hbar \lambda(1-d)$
 is generally identified with the inverse of the classical escape time
 of the system.  We finally caution that the decay rate will be
 subject to sources of \textit{fluctuations} which are beyond the
 scope of our analysis. Notably, the Lyapunov exponents may vary
 between trajectories, and along individual trajectories. The escape
 from the repeller may introduce additional uncertainty.  Our result,
 thus, yields a characteristic value for the decay rate, where the
 important role of fluctuations is left unaccounted for.  Other
 effects not captured by our analysis include transient features of
 the classical dynamics \textit{outside} the repeller's area which, as
 recent work shows~\cite{ermann, carlo}, may have important influence
 on the resonances of open quantum system.

\subsection{Fractal Weyl law and random matrix regime} 
For completeness we note that a finite quantum mechanical decay rate
is attributed to states located in the vicinity of exceptionally long
trajectory. Due to the exponential scarcity of these trajectories, the
corresponding phase space measure scales as~\cite{Schomerus} $n_\Gamma
= \Omega_E e^{-t_E/t_d}\sim \Omega_E \hbar^{1/t_d \lambda}$, where
$\Omega_E$ is the phase space volume of the energy shell (in units
$\hbar$), implying that $n_\Gamma \sim \hbar^{-d_f}$ with
\textit{fractal} dimension $d_f = 1-1/\lambda t_d$~\cite{fnp}. We
finally note that in the quantum regime, $t_E < t_d$, which is
complementary to the semiclassical regime $t_d<t_E$ studied here,
random matrix scattering theory predicts~\cite{SchomerusRev} $\Gamma
\sim {\hbar \over t_d}\ln (\tilde{E}t_d/\hbar)$, where $\tilde{E}$ is
some cut-off energy scale. Comparison with \eqref{eq:1} shows that the
two results match at the boundary $t_d \sim t_E$. However, we cannot
say whether this matching is coincidental or not.

\section{Summary}
\label{sec:summary} 
We have formulated a semiclassical theory of quantum escape processes
in open chaotic systems.  The most important single contribution of
our approach is that it quantitatively describes how deterministic
escape after the traversal of generic short trajectories through the
system gives way to \textit{quantum mechanical decay} on long
trajectories. The latter define the support of resonances whose life
times we estimated by imposing effective phase space boundary
conditions.  Somewhat counterintuitively, it turns out that the
ensuing decay rates are classically short $\Gamma_0\sim \hbar
\lambda$, although the relevant escape dynamics takes place on
long trajectories $T>t_E$. Finally, the escape of individual long
trajectories as described in the present paper defines only an
initializing stage of the decay of a more complicated repeller
structure. As a result, the decay rate $\Gamma$ is subject to
renormalization $\Gamma_0\to \Gamma=\hbar \lambda (1-d)$ where $d$ is
the fractal repeller dimension. Qualitatively, the renormalization
factor accounts for the probability that a state gets re-captured by
the repeller structure after escaping an individual
trajectory. However, a quantitative description of that secondary
mechanism is beyond the scope of our approach.

\acknowledgements We thank P.~W.~Brouwer, S.~Nonnenmacher and 
H. Schomerus for
discussions and M.~Novaes for pointing out an inconsistency in the
interpretation of our results as given in the first version of this paper. 
Work supported by SFB/TR 12 of the Deutsche Forschungsgemeinschaft.

\begin{appendix}

\section{Regularization}

We here discuss how quantum fluctuations regularize the unlimited
classical contraction of the  stable coordinate $s$ in a system
with globally hyperbolic dynamics. In the language of Eqs.~(6)
and~(7) of the main text, the dynamics of the variable $s$ is
described by a differential equation of the structure
\begin{align}
\label{sdgl}
\left( s \partial_s + \sum_{n\geq1} c_n \hbar^{2n+1} \partial^{2n+1}_s \right) 
f(s) = - \alpha f(s)
\end{align} 
where $\alpha>0$, and in a manner inessential to the present argument
the coefficients $c_n$ may depend on the variables
$q,u,s$. 

Considering positive starting values, $s>0$ (the extension to negative
values is straightforward), we introduce a Laplace representation
\begin{align}
\label{slt}
f(s)=\int_0^\infty dz\, e^{-sz} g(z)
\end{align}
in which \eqref{sdgl} takes the form~\cite{sfn}
\begin{align}
\label{sLTeq}
\partial_z  g(z)=  
- \left( {1 - \alpha  \over z}  +  \sum_{n\geq1} c_n \hbar^{2n+1} z^{2n}  \right) g(z),
\end{align} 
The general solution of this equation is found by straightforward
integration over $z$, and when inserted into \eqref{slt}
gives 
\begin{align}
\label{sgs}
f(s)= c_0\int_0^\infty dz\, e^{-sz}
  z^{\alpha-1} e^{-\sum_{n\geq 1} {c_n\over 2n+1} (\hbar z)^{2n+1}}
\end{align}
with an integration constant $c_0$.
Eq.~\eqref{sgs} now illustrates the role played by higher differential operators
in \eqref{sdgl}.

To make the point, let us for the moment consider the first order differential equation 
obtained from \eqref{sdgl} by setting all $c_n=0$. The resulting
function 
\begin{align}
\label{sgs0}
f^0(s)= c_0\int_0^\infty dz\, e^{-sz}
  z^{\alpha-1}
  ={c_0\over s^\alpha}
  \end{align} 
then displays the singular at small values of $s$ plaguing the
classical evolution equation of the stable coordinate. 

In the full solution Eq.~\eqref{sgs} the exponential factor 
 $e^{-\sum_{n\geq 1} {c_n\over 2n+1} (\hbar z)^{2n+1}}$ cuts the small
 $s$/large $z$ singularity at values $z\sim 1/\hbar$. The resulting
 integral can be estimated by a regularized function
\begin{align}
\label{sreg}
f(s)
={ c_0  \over (s + \hbar)^\alpha}.
\end{align} 
Finally notice that our argument crucially relies on assumed
positivity of the coefficients $c_n$. While the present construction
cannot prove this feature, positivity is required on principal grounds
to ensure stability of the dynamics. (Otherwise the Wigner
distribution would cease to exist.) To actually demonstrate this
stability, one has to work harder as in,
e.g., Refs.~\cite{ZurekPaz,sDicke}. A discussion
tailored to the present formalism is contained in Ref.~[\onlinecite{AndreevBilliards}].

\end{appendix}


\begin{thebibliography}{99}
\bibitem{fnp}
Here we used $\Omega_E = L^2 p/\hbar \ell$,
where  $L$ sets the size of cavity cavity and 
$\hbar/\ell$ the characteristic momentum resolution.
\bibitem{sfn} 
For the sake of the argument, we focus on solutions subject to the
condition
$[zg(z)]_{z=0}=0$, such that \eqref{sLTeq} does not contain a boundary term.
\end{thebibliography}
\end{document}